\documentclass[prl,aps,twocolumn,superscriptaddress,10pt,showpacs]{revtex4-1}
\usepackage{graphicx}
\usepackage{epstopdf}
\usepackage{amsmath,amssymb,dsfont,upgreek}
\usepackage{bm}
\usepackage{natbib}
\usepackage[colorlinks=true,  linkcolor=blue, citecolor=blue, urlcolor=blue,hyperfootnotes=false]{hyperref}
\usepackage{soul}

\begin{document}
\date{\today}

\title{Temperature-induced spontaneous time-reversal symmetry breaking on the honeycomb lattice.}

\author{Wei Liu}
\affiliation{Physics Department, City College of the City University of New York, New York, NY 10031, USA}
\author{Alexander Punnoose}%
\email{punnoose@sci.ccny.cuny.edu}
\affiliation{Physics Department, City College of the City University of New York, New York, NY 10031, USA}
\affiliation{Instituto  de F\'\i sica Te\'orica $-$ Universidade Estadual Paulista, R.\ Dr.\ Bento Teobaldo Ferraz 271, Barra Funda, S\~ao Paulo - SP, 01140-070, Brazil}

\begin{abstract}
Phase transitions  involving spontaneous  time-reversal symmetry breaking  are studied  on the honeycomb lattice at finite hole-doping with next-nearest-neighbor repulsion. We derive an exact expression for the mean-field equation of state in closed form, valid at temperatures much less than the Fermi energy.  Contrary to standard expectations, we find  that thermally induced intraband particle-hole excitations can create and stabilize a uniform metallic phase with broken time-reversal symmetry  as the temperature is \textit{raised}  in a region where  the groundstate is  a trivial metal.

\end{abstract}


\maketitle

\textit{Introduction.}\textemdash\; A popular proposal to  break time-reversal ($\mathcal{T}$) symmetry in an electronic system  without invoking the spin degrees of freedom and without breaking translational symmetry is to  imagine microscopic electronic current loops   arranged so that their net moment vanishes   in each unit cell \cite{haldane_graphene_TRV}. %
In multiband systems,   $\mathcal{T}$ symmetry breaking is intimately related to  the  condensation of  interband particle-hole pairs when the electron-electron scattering is strong enough \cite{[{Particle-hole condensation in a strongly interacting system when the bandgap $E_G\geq 0$ is  distinct from the  instability predicted to occur in a semimetal with  $E_G<0$ in which a finite density of \textit{real} particles and holes exist in equilibrium in the non-interacting limit;  the Hamiltonian in the latter case maps to   the  BCS (Bardeen-Cooper-Schrieffer) model in a Zeeman field and has been studied extensively in the past: }]Mott_Transition,*knox,*Cloizeaux,*Keldysh_Kopaev, *Kozlov_Maksimov, *jerome_rice_kohn, *RMP_halperin_rice, *Nozieres_Comte}.
The reduced symmetry of the vertex function $\mathcal{D}_{\bm{k}}$ that couples the interband bilinear operators $a_{\bm{k}}^\dagger b_{\bm{k}}$   determines both  the symmetry breaking and  the symmetry of the order parameter $\varphi\sim\sum_{\bm{k}} \mathcal{D}_{\bm{k}}\langle a_{\bm{k}}^\dagger b_{\bm{k}}\rangle$. In particular, $\mathcal{T}$ symmetry will be  broken if $\mathcal{TD}_{\bm{k}}\mathcal{T}^{-1}
= \mathcal{D}^*_{-\bm{k}}\neq  \mathcal{D}_{\bm{k}}$   \cite{varma_1997, fradkin_sun_TRV,punnoose_TRV_XL}. 

Metallic systems with non-trivial topological properties, generically referred to as topological Fermi liquids (TFLs) \cite{haldane_TFL}, have come to  play an important role in the study of marginal and other non-Fermi liquid phenomena \cite{varma_SFL_review}.  Various instances  \cite{honeycomb_spain_PRL,Zhang_TMI,QBCP_fradkin,honeycomb_Weeks_Franz,honeycomb_fiete,triangle_tieleman,imada_gapless,honeycomb_spain_PRB,honeycomb_herbut} of TFLs and topological Mott insulators (TMIs) at commensurate fillings have been  realized on the honeycomb  lattice using a combination of mean-field theory and numerics.  %


Although the breaking of $\mathcal{T}$ symmetry opens a gap at the Dirac points \cite{Zhang_TMI}, no gap opens at the Fermi surface in mean-field theory \cite{punnoose_TRV_XL}. Thus a continuum of gapless intraband particle-hole excitations exists in the ordered phase in one-to-one correspondence with the free Fermi gas. They provide the leading temperature corrections to the free energy $\Omega_\textrm{MF}(T,\mu,\lambda)$ at the mean-field level. Here, $\mu$ is the chemical potential, $\lambda$ is the external field that couples to the order parameter, and $T$ is temperature.
The order parameter $\varphi$ is given by the relation $\partial\Omega_\textrm{MF}/\partial\lambda=-\varphi$; the resulting integral equation is in general not restricted to the vicinity of the Fermi surface but extends over a large portion of the Fermi sea up to a cut-off scale $\Lambda\sim 1/a$, where $a$ is some microscopic scale determined by the range of the interaction. This results in an increased sensitivity to the  number of particles $N$ within the cut-off energy.  Thus the equation for the order parameter has to be solved simultaneously with the number equation $\partial\Omega_\textrm{MF}/\partial\mu=-N$. The solutions, if derivable, can be used to construct the canonical free energy $F_\textrm{MF}(T,N,\varphi)=\Omega_\textrm{MF}(T,\mu,\lambda)+\lambda \varphi +\mu N$.

In this paper, we obtain an analytical solution  for the mean-field  equation of state $\partial F_\textrm{MF}/\partial \varphi = \lambda$ and $\partial F_\textrm{MF}/\partial N = \mu$   for $N$ interacting spinless particles  with next-nearest-neighbor repulsion  on the honeycomb lattice away from half-filling.  
The solutions are valid in the  limit $T\ll \epsilon_F$, where $\epsilon_F$ is the Fermi energy; physically, in this  limit only the  \textit{intraband} particle-hole excitations around the Fermi surface provide a significant contribution to the partition function at finite $T$, while  the pair-breaking \textit{interband} excitations can be neglected.   

Analysis of these equations in the limit $\lambda\rightarrow 0$ allows us to  identify and study both the quantum critical point (QCP) and the low temperature phase diagram. 
We find that the thermal smearing of the Fermi surface has the curious effect of favoring the condensation of  interband particle-hole pairs resulting in a finite temperature phase transition into an ordered state with broken $\mathcal{T}$ symmetry. This leads us to the novel conclusion that a stable topological state can be created by raising the temperature.

\textit{Model.}\textemdash\; We consider the extended Hubbard model of spinless electrons on a honeycomb lattice with nearest-neighbor ($nn$) hopping and next-nearest-neighbor ($nnn$) repulsion.  
It was concluded in Ref. [\onlinecite{honeycomb_spain_PRB}], after extensive mean-field studies of this model at $T=0$,  that a finite $nnn$ repulsion is necessary to stabilize a $\mathcal{T}$ symmetry broken metallic phase in the vicinity of half-filling, which is consistent with Ref. [\onlinecite{Zhang_TMI}] where the half-filled case was first studied. They compete with other charge-modulated states that occupy most of the parameter space when $nn$ repulsion is included. Here, we  isolate and study the stability of the $\mathcal{T}$ broken phase at finite $T$, keeping only the $nnn$ repulsion while neglecting the $nn$ repulsion.

For the $C_{6v}$ point symmetry group of the underlying honeycomb lattice, only the one-dimensional irrep (called $B_1$ \cite{tinkham_book}) with a reduced $C_6$ symmetry%
\begin{equation}
\mathcal{D}_{\bm{k}}=\frac{1}{\sqrt{3}}\left[\sin(\bm{k\cdot t}_1)+\sin(\bm{k\cdot t}_2)+\sin(\bm{k\cdot t}_3)\right]
\label{eqn:Dk}
\end{equation}
is real and odd so that  $\mathcal{D}^*_{-\bm{k}}\neq \mathcal{D}_{\bm{k}}$ \cite{punnoose_TRV_XL}.  (The vectors $\bm{t}_l$, with $l=1,2,$ and $3$, are the  basis vectors of the hexagonal Bravais lattice.) 
The effective Hamiltonian that isolates this irrep in the particle-hole sector reduces to \cite{punnoose_TRV_XL}
%
\begin{equation}
K=\sum_{\bm{k}}\psi^\dagger_{\bm{k}}\left[\epsilon_{\bm{k}}
{\uptau}_3-\mu\right]\psi_{\bm{k}} -\frac{V}{2L^2}\sum_{\bm{q}}\hat{\Phi}^\dagger(\bm{q})\hat{\Phi}(\bm{q})
\label{eqn:K}
\end{equation}
The basis $\psi^\dagger_{\bm{k}}=(b^\dagger_{\bm{k}},a^\dagger_{\bm{k}})/\sqrt{2}$ diagonalizes the kinetic term after a unitary rotation
$a_{\bm{k}}=e^{-\frac{i}{2}\theta_{\bm{k}}}A_{\bm{k}}+e^{\frac{i}{2}\theta_{\bm{k}}}B_{\bm{k}}$ and $b_{\bm{k}}=e^{-\frac{i}{2}\theta_{\bm{k}}}A_{\bm{k}}-e^{\frac{i}{2}\theta_{\bm{k}}}B_{\bm{k}}$, where
$A_{\bm{k}}$ and $B_{\bm{k}}$ are the momentum space representations of the on-site  annihilation operators on the two sub-lattices of the honeycomb lattice. The  $\uptau_i$'s are Pauli matrices that act on the band indices.  The angles are obtained from $\epsilon_{\bm{k}}e^{i\theta_{\bm{k}}}=\sum_l\exp(i\bm{k\cdot d}_l)$, where the vectors $\bm{d}_l$  connect  the $nn$ atoms on the two sub-lattices. The band energies $\epsilon_{\bm{k}}=\sqrt{3+2C_{\bm{k}}}$, where $C_{\bm{k}}=\sum_l\cos(\bm{k\cdot t}_l)$, have vanishing gaps at two inequivalent Dirac points, $\pm \bm{k}_D$.   All energies are measured from the Dirac points in units of the  hopping integral $t$.

The bilinear  operator  in the interaction term reads   \cite{punnoose_TRV_XL}
\begin{equation}
\hat{\Phi}(\bm{q})=\sum_{\bm{k}}\mathcal{D}_{\bm{k}}\psi^\dagger_{\bm{k+\frac{q}{2}}}\left[e^{\frac{i}{2}\left(\theta_{\bm{k+\frac{q}{2}}}-\theta_{\bm{k-\frac{q}{2}}}\right)\uptau_1}\uptau_1\right]\psi_{\bm{k-\frac{q}{2}}}
\label{eqn:Phi_q}
\end{equation}
 Although our choice of the $\psi$ basis  complicates $\hat{\Phi}$ by introducing additional  phases, defining the order parameter as $\varphi\sim \langle \hat{\Phi}(\bm{q}=0)\rangle$ makes explicit the mapping of $K$ in (\ref{eqn:K})  to the transverse-field Ising model:  
The system undergoes a transition at a critical $V_c$ from a ``paramagnet"  with all spins pointing along $\uptau_3=-1$  (fully occupied lower band when $\varphi=0$)   to a correlated  ``Ising ferromagnet"  with the spins pointing along $\uptau_1$ when $\varphi\neq 0$.
It is well known that the collective excitations of the transverse-field Ising model are gapped \cite{deGennes_Ising,brout_Ising}. However, in a metallic system the low-energy fermionic excitations qualitatively changes the nature of the transition \cite{imada_gapless,imada_review,punnoose_TRV_XL}. This is the key point that we analyze here.

\textit{Fixed $\mu$.\,}\textemdash\;  The action corresponding to  $K_\lambda=K-\lambda \hat{\Phi}(\bm{q}=0)$, where the Hamiltonian $K$ is defined in (\ref{eqn:K}) and $\lambda$ is a constant source term that couples to the field operator $\hat{\Phi}(\bm{q}=0)=\sum_{\bm{k}}\mathcal{D}_{\bm{k}}\psi^\dagger_{\bm{k}}\uptau_1\psi_{\bm{k}}$ ,  is derived the standard way  \cite{book_Altland_Simons}  by first decoupling the quartic fermion interaction terms using   Hubbard-Stratonovich (HS) fields.
Since  $\hat{\Phi}$  is Hermitian, the  HS fields, $\phi_q$, are real, i.e., $\phi^*_q=\phi_{-q}$; note that, when $\bm{q}=0$, the phase factors in (\ref{eqn:Phi_q}) drop out leading to considerable simplifications.

The HS transformation allows the fermions to be formally integrated out. The resulting partition function is written as  $Z=\int D[\phi]\exp(-S)$, where the bosonic action \cite{punnoose_TRV_XL} 
$S= ({L^2}/{2\beta V})\sum_q \phi_{-q}\phi_q-\textrm{Tr}\ln G^{-1}$ with 
$G^{-1}_{k,k'}=\left(-i\epsilon_n - \mu +\epsilon_ {\bm{k}}\uptau_3-\lambda\mathcal{D}_{\bm{k}}\uptau_1\right)\delta_{k,k'}
-T\phi_{k-k'}\mathcal{D}_{\bm{k}_+}\hat{f}_{\bm{k,k'}}$.
The shorthand notations, $k\equiv( \bm{k},ik_n)$ and $k-k' = q \equiv ( \bm{q},iq_m)$, where $k_n$ $  (q_m)$ are the odd (even) Matsubara frequencies, are used; the Fourier transform  is defined as $\phi(x)=T\sum_m{L^{-2}}\sum_ {\bm{q}} e^{-iq_m\tau + i \bm{q}\cdot  \bm{r}} \phi_q.$  The operator $\hat{f}_{\bm{k,k}'}$ represents all the terms within the square brackets in (\ref{eqn:Phi_q}); note $\hat{f}_{\bm{k,k}}=\uptau_1$. The  momentum  $\bm{k}_+=(\bm{k}+\bm{k}')/2$ in $\mathcal{D}_{\bm{k}_+}$.

Assuming a uniform solution  $\phi_q=\beta\delta_{m,0}\delta_{\bm{q},0}\varphi$ in $S$, followed by a few standard manipulations involving the identity $\textrm{Tr}\ln \hat{A}=\ln \det{\hat{A}}$ and summing over the  frequencies  \cite{book_Altland_Simons,bluebook}, the action after scaling $S=(\beta L^2)\mathcal{L}$ can be written as $\mathcal{L}={\varphi^2}/{(2V)} +\mathcal{L}_{+} + \mathcal{L}_{-}$, where \cite{imada_gapless}
\begin{equation}
\mathcal{L}_{\pm}=- \frac{T}{L^2}\sum_{\bm{k}}\ln\left(1+e^{-\mathcal{E}_{\lambda,k}^\pm/T}\right)
\label{eqn:Omega_pm}
\end{equation}
The renormalized bands equal $\mathcal{E}_{\lambda,k}^\pm=\pm E_{\lambda,k} -\mu$, where  $E_{\lambda,k}=({\epsilon_{\bm{k}}^2+\mathcal{D}_{\bm{k}}^2\varphi_\lambda^2})^{1/2}$,   $\varphi_\lambda=\varphi+\lambda$ is the shifted order parameter and the bare energies  $\pm\epsilon_{\bm{k}}$ are defined in (\ref{eqn:K}). 
%
%
From here on, we confine ourselves to the linear regime around the Dirac points. As is well known,  the bare spectrum is linear
$\epsilon_{\bm{p}}= \alpha |\bm{p}|$ to first order in $\bm{p}=\bm{k}-\bm{k}_D$
with $\alpha =3/2$.
Furthermore, we assume  $|\mathcal{D}_{\bm{k}}|=|\mathcal{D}_{\bm{k}_D}|$;
note that coincidentally the constant $|\mathcal{D}_{\bm{k}_D}|=3/2=\alpha$, as a result the band energies assume the simpler form
\begin{equation}
E_{\lambda,p}=\sqrt{\epsilon_p^2+\mathcal{D}_{\bm{k}_D}^2\varphi_\lambda^2}=\alpha\sqrt{p^2+\varphi_\lambda^2}
\label{eqn:Ep}
\end{equation}

The linear approximation requires a cut-off momentum $p_\Lambda$, which we determine by fixing the number of electrons in the cones at $T=0$. It is easily seen that the mean-field Hamiltonian  $[H_\textrm{MF},\hat{n}_{\bm{k}}]=0$ conserves the particle number in each $\bm{k}$ \cite{punnoose_TRV_XL}, hence  the number of electrons $N_e$ is always equal to the number of states between the non-interacting cut-offs $p_F$ and $p_\Lambda$  independently of $\varphi$, i.e.,
\begin{equation}
n_e=\frac{N_e}{L^2}=\frac{2}{L^2}\sum_{\bm{k}}=2\int_{p_F}^{p_\Lambda}\frac{pdp}{2\pi}=\frac{1}{2\pi}(p_\Lambda^2-p_F^2)
\label{eqn:ne_0a}
\end{equation}
The factor $2$ accounts for the two Dirac cones. For definiteness, we consider  hole doping; the hole density $n_h=n-n_e$, where $n=\epsilon_\Lambda^2/(2\pi\alpha^2)$ and  $n_h=\epsilon_F^2/(2\pi\alpha^2)$, and the bare energies $\epsilon_\Lambda=\alpha p_\Lambda$ and $\epsilon_F=\alpha p_F$.  It is convenient to  write the renormalized cut-offs in terms of the  rescaled densities $\tilde{n}=(\pi\alpha^2)n$ and $\tilde{n}_e=(\pi\alpha^2)n_e$ as  $\Lambda_\lambda=\sqrt{\epsilon_\Lambda^2+\Delta_\lambda^2}=\sqrt{2\tilde{n}+\Delta_\lambda^2}$ and $E_{\lambda,F}=\sqrt{\epsilon_F^2+\Delta_\lambda^2}=\sqrt{2\tilde{n}_h+\Delta_\lambda^2}$.  It follows that  $2|\Delta_\lambda|=2\alpha|\varphi_\lambda|$ is the gap at the Dirac points.
Eq.  (\ref{eqn:ne_0a}) then reduces to
\begin{equation}
\tilde{n}_e=\tilde{n}-\tilde{n}_h=\frac{1}{2}(\Lambda_\lambda^2-E_{\lambda,F}^2)
\label{eqn:ne_0}
\end{equation}
which is independent of $\Delta_\lambda$ as discussed. 

The terms $\mathcal{L}_\pm$ defined in Eq. (\ref{eqn:Omega_pm}) are valid for all temperatures and fillings.
Here, we restrict ourselves to the temperature range $T/E_{\lambda,F}\ll 1$, which corresponds to 
suppressing all terms 
that fall off exponentially as $e^{-\epsilon_F/T}$ and higher. (This approximation breaks down at half-filling, where $\mu=0$ from particle-hole symmetry. For a detailed analysis of the half-filled case, see \footnote{
The TMI has been analyzed in detail in Ref. [\onlinecite{imada_gapless}]; the deviation from the standard LGW theory, peculiar to the half-filled case at $T=0$, vanish both at finite $T$ and/or doping. For the interested reader, the relationship between the parameters in   [\onlinecite{imada_gapless}] with those in this paper are:  $\zeta=\varphi/V,\, \Lambda=p_\Lambda,\, t=\alpha/\sqrt{3}$ and $ V_2/t=V/3$.
}.) 

To this end, we first evaluate the momentum sums in (\ref{eqn:Omega_pm}), which can be done analytically, and then expanded using the asymptotic properties of the polylogarithm functions \cite{book_functions}. The result of the expansion when combined with the $\varphi^2/(2V)$ term in $\mathcal{L}$ gives (see appendix [A1] for the derivation of Eq. (\ref{eqn:Omega_lowT}))
\begin{equation}
\tilde{\mathcal{L}}=\frac{\pi\Delta^2}{2V}-\frac{\Lambda_\lambda^3}{3}+\frac{\Lambda_\lambda^2|\mu|}{2}-\frac{|\mu|^3}{6}-\frac{\pi^2T^2|\mu|}{6}
\label{eqn:Omega_lowT}
\end{equation}
Here, $\tilde{\mathcal{L}}=(\pi\alpha^2)\mathcal{L}$ (similar to the rescaling of $\tilde{n}$). Since hole doping is assumed,  $\mu=-|\mu|\neq 0$.  
%
The $T^2$ contribution is the familiar Sommerfeld  correction that originates from the particle-hole excitations at the Fermi surface; it plays a key role in what follows. 

The mean-field  free energy $\Omega(\mu,\lambda)$ is defined as  %
$\Omega(\mu,\lambda)=\tilde{\mathcal{L}}(\bar{\Delta},\mu,\lambda)$,
where  $\bar{\Delta}$ satisfies the saddle-point equation ${\partial\tilde{\mathcal{L}}}/{\partial \Delta}|_{\bar{\Delta}}=0$ \cite{book_negele_orland}; differentiating (\ref{eqn:Omega_lowT}), we get
\begin{equation}
\frac{\pi\bar{\Delta}}{V}= \bar{\Delta}_\lambda\left(\bar{\Lambda}_\lambda-|\mu|\right)
\label{eqn:saddle_point_lambda}
\end{equation}
(The bars denote that the quantities are evaluated at the saddle-point.) 
The same equation is obtained by setting the total derivative $d\tilde{\mathcal{L}}/ d\lambda=-(\pi \alpha^2/L^2)\langle \hat{\Phi}(\bm{q}=0)\rangle=-\alpha(\pi\bar{\Delta}/V)$ at the saddle-point, which is consistent with assigning  $\bar{\Delta}$ as the order parameter in the limit $\lambda\rightarrow 0$.

\textit{Fixed $n_e$.\,}\textemdash\;
Since $\Omega$ as defined  is the free energy, the particle density corresponds to $\tilde{n}_e=\partial{\Omega}/\partial|\mu|$; differentiating (\ref{eqn:Omega_lowT}) again, we get
$\tilde{n}_e=(\bar{\Lambda}_\lambda^2-\mu^2)/2-{\pi^2T^2}/{6}$. 
Comparing $\tilde{n}_e$  with its $T=0$ value  in (\ref{eqn:ne_0}), we obtain %
\begin{equation}
\mu(T)^2=\bar{E}_{\lambda,F}^2-\frac{\pi^2}{3}T^2=2\tilde{n}_h+\bar{\Delta}_\lambda^2-\frac{\pi^2T^2}{3}
\label{eqn:mu_lowT}
\end{equation}

Solving Eqs. (\ref{eqn:saddle_point_lambda}) and (\ref{eqn:mu_lowT})  for $\lambda$ and $\mu$ in terms of $\bar{\Delta}$ and $\tilde{n}_e$ and substituting the solutions  in
\begin{equation}
\mathcal{F}(\bar{\Delta},\tilde{n}_e)=\Omega(\mu,\lambda)+\alpha\lambda\frac{\pi\bar{\Delta}}{V}+\mu \tilde{n}_e
\label{eqn:F_def}
\end{equation}
gives the free energy $\mathcal{F}(\bar{\Delta},\tilde{n}_e)$ (per unit volume).

\begin{widetext}
Fortunately, it is possible to solve (\ref{eqn:saddle_point_lambda}) and (\ref{eqn:mu_lowT}) in closed form. For the equation of state, we get (See appendix [A2] for the derivation of Eq. (\ref{eqn:sol_lam}))
\begin{equation}
\alpha\lambda=\left(\frac{\pi\bar{\Delta}}{V}\right)\left[\left(
\frac{(2\tilde{n}+2\tilde{n}_h-\pi^2T^2/3)+2\sqrt{2\tilde{n}(2\tilde{n}_h-\pi^2T^2/3)+(\pi\bar{\Delta}/V)^2}}
{(2\tilde{n}-2\tilde{n}_h+\pi^2T^2/3)^2-4(\pi\bar{\Delta}/V)^2}\right)^{1/2}-\frac{V}{\pi}\right]
\label{eqn:sol_lam}
\end{equation}
Substituting Eq. (\ref{eqn:sol_lam})  back into (\ref{eqn:mu_lowT}) gives the solution for $|\mu|$. Note that the cut-offs are absorbed in $\tilde{n}$ and $\tilde{n}_h$. 
\end{widetext}
%
%
%

At $T=0$, 
the QCP, $V_c$, is obtained by setting $\bar{\Delta}$ and $\lambda=0$ in (\ref{eqn:sol_lam}). This reproduces the result   \cite{punnoose_TRV_XL}
\begin{equation}
V_c=\frac{\pi}{\sqrt{2\tilde{n}}-\sqrt{2\tilde{n}_h}}=\frac{\pi}{\epsilon_\Lambda-\epsilon_F}.
\end{equation}
Our key observation is that when $T>0$, a non-trivial solution emerges   in the region $V<V_c$ at a finite temperature $T_c$.  To see this, we set  $\bar{\Delta}=0$ in (\ref{eqn:sol_lam}) and get
\begin{equation}
V(T_c)=\frac{\pi}{\sqrt{2\tilde{n}}-\sqrt{2\tilde{n}_h-\pi^2T_c^2/3}}<V_c
\label{eqn:Vc_T}
\end{equation}
Eq. (\ref{eqn:Vc_T}) can be inverted to give the finite temperature phase boundary. For $V$ close to $V_c$, we get
\begin{equation}
T_c(V)=\sqrt{\frac{6\epsilon_F}{\pi V_c}}\times \sqrt{\frac{V_c-V}{V_c}} \hspace{0.25cm}\textrm{as}\hspace{0.25cm}V\rightarrow V_c^-
\label{eqn:Tc}
\end{equation}
This is an intriguing result: contrary to standard expectations,
a topologically non-trivial metallic phase with broken $\mathcal{T}$ symmetry, which is unstable at $T=0$ for $V<V_c$, is stabilized as the temperature is raised above $T_c$. The prefactor $\sqrt{\epsilon_F}$ is a reminder that the $T^2$ term originates from the  intraband particle-hole excitations; hence, this thermally induced  finite temperature $\mathcal{T}$ symmetry broken phase only exists at finite doping.

It is easily verified that  Eq. (\ref{eqn:Vc_T}) can be obtained directly from (\ref{eqn:saddle_point_lambda}) and (\ref{eqn:mu_lowT}), instead of using the solution in (\ref{eqn:sol_lam}). Neither way, however,  guarantees that the $\bar{\Delta}\neq 0$ phase is stable and is lower in free energy than the trivial phase with $\bar{\Delta}=0$. For this,  the difference in the free energies $\delta \mathcal{F}(\bar{\Delta})=\mathcal{F}(\bar{\Delta},\tilde{n}_e)-\mathcal{F}(0,\tilde{n}_e)$ has to be analyzed.

\begin{figure}[t]
\center{\includegraphics[width=0.9\linewidth]{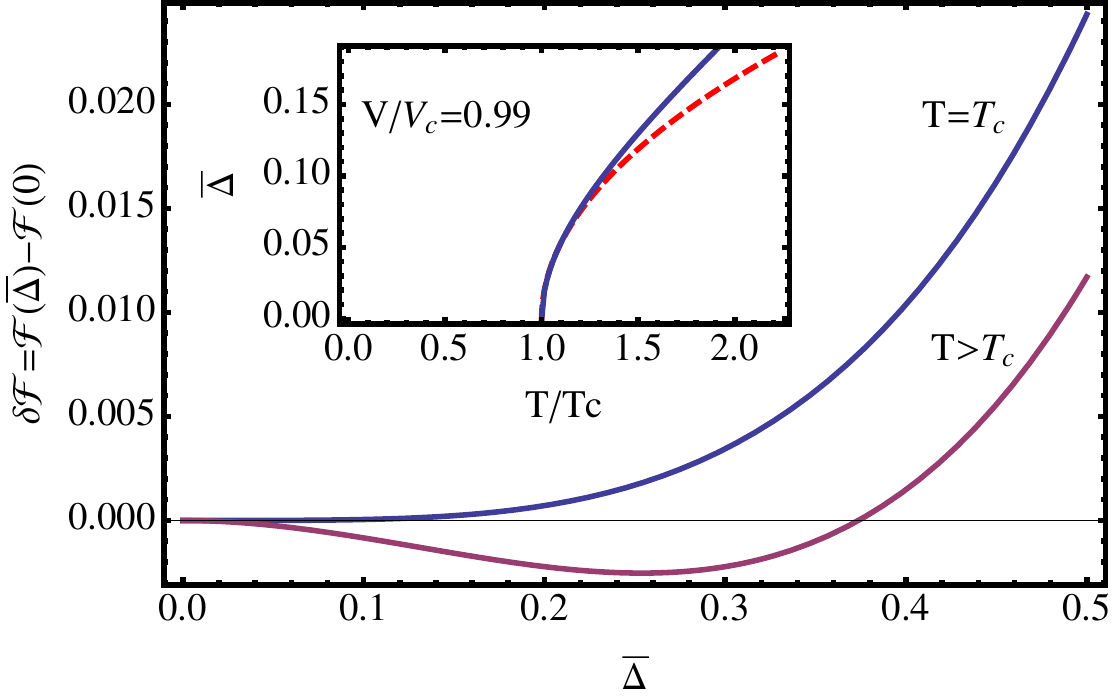}}
\caption{The difference of the canonical free energy $\delta \mathcal{F}(\bar{\Delta})$ defined in (\ref{eqn:F_def}) is evaluated numerically  keeping the density $\tilde{n}_e$ fixed and plotted  for $V<V_c$ as a function of $\bar{\Delta}$ for $T=T_c$ and $T>T_c$.  A clear minimum is observed as the temperature is raised above $T_c$  (see  Eq. (\ref{eqn:Tc})), signifying the formation of a topological liquid phase at finite $T$. The evolution of the order parameter $\bar{\Delta}(T)$ (the location of the minimum) is shown in the inset by the solid line; the dashed line is the approximate expression for $\bar{\Delta}$ derived in (\ref{eqn:dminus}). The filling is fixed at $\epsilon_F/\epsilon_\Lambda=\sqrt{n_h/n}=0.1$ and $V_c=1.3$.
}
\label{fig:fig}
\end{figure}

Although it is straightforward to calculate $\delta \mathcal{F}$ given the equation of state (\ref{eqn:sol_lam}), the algebra is tedious and the result  in its full generality is not very informative. The fact that the state is stable can easily be seen by plotting it numerically as  shown in Fig. \ref{fig:fig}. 
A clear minimum is observed  when $T>T_c$ in the region where $V<V_c$.  

The curvature at the minimum can be obtained analytically by differentiating  the equation of state in (\ref{eqn:sol_lam}), $(\alpha\pi/V)\lambda=\partial\mathcal{F}/\partial \bar{\Delta}|_{\tilde{n}_e}$, holding $\tilde{n}_e$ fixed:
\begin{equation}
\left.\frac{\partial^2\mathcal{F}}{\partial\bar{\Delta}^2}\right|_{\tilde{n}_e}
=\left.\frac{\alpha\pi}{V}\frac{\partial\lambda}{\partial\bar{\Delta}}\right|_{\tilde{n}_e}
\approx\frac{\alpha\bar{\Delta}^2}{\epsilon_\Lambda\epsilon_F} >0 \hspace{0.25cm}\textrm{as}\hspace{0.25cm}T\rightarrow T_c
\label{eqn:stability}
\end{equation}
The derivative is evaluated at $V=V(T_c^+)$, slightly above the critical temperature where $\bar{\Delta}$ is small but finite. 

Approximate analytic expressions for the variation of the  order parameter $\bar{\Delta}$   as a function of $V$ and $T$ when $\bar{\Delta}^2/(\epsilon_F\epsilon_\Lambda)\ll 1$ are easily obtained by expanding  (\ref{eqn:sol_lam}) near $V_c$.
Setting $\lambda=0$ and expanding (\ref{eqn:sol_lam}), we get
%
\begin{equation}
\frac{\bar{\Delta}^2}{\epsilon_\Lambda\epsilon_F}=\frac{\epsilon_F(T)}{\epsilon_F}\times \frac{V^2}{V(T)^2}\times \left(\frac{V^2-V(T)^2}{V(T)^2}\right)
\label{eqn:delta2}
\end{equation}
Here $\epsilon_F(T)=\sqrt{\epsilon_F^2-\pi^2T^2/3}$ and the function $V(T)$ has the same form as $V(T_c)$ in (\ref{eqn:Vc_T}) with $T_c$ replaced by $T$.  
Near $T_c$,  $V(T)$  can be expanded to linear order in $(T_c/\epsilon_F)$ in powers of $\delta T=T-T_c$ as
\begin{equation}
\frac{V(T_c)-V(T)}{V(T_c)^2}\approx \frac{\pi}{3}\left[\left(\frac{T_c}{\epsilon_F}\right)\delta T+\frac{(\delta T)^2}{2\epsilon_F}\right]
\label{eqn:V2}
\end{equation}
From  (\ref{eqn:Vc_T}), we see that   $T_c\rightarrow 0$ as $V(T_c)\rightarrow V_c$, hence $T_c\neq 0$ only in the weak-coupling regime $V<V_c$. The  $(\delta T)^2$ term is therefore necessary in the strong-coupling regime  $V>V_c$ as  the linear term vanishes there.

The behavior of $\bar{\Delta}$ as $V$ crosses from $V_c^-$ to $V_c^+$ is analyzed below using Eqs. (\ref{eqn:delta2}) and (\ref{eqn:V2}): 

(\textit{i}) For $V<V_c$, we fix the potential  at $V=V(T_c)$ in (\ref{eqn:delta2}) and expand $V(T)=V(T_c+\delta T)$ using (\ref{eqn:V2}) to leading order in $\delta T/T_c$ and get
\begin{equation}
\bar{\Delta}(T)^2=\frac{2\pi^2T_c^2}{3}\left( \frac{T-T_c}{T_c}\right)  \hspace{0.25cm}\textrm{as}\hspace{0.25cm}V\rightarrow V_c^-
\label{eqn:dminus}
\end{equation}
A plot of this function indicated by dashed lines in the inset to Fig. \ref{fig:fig} agrees accurately with the numerically extracted minimum (solid line in the inset) as $T\rightarrow T_c^+$. 
%

(\textit{ii}) For $V>V_c$, we set $T_c=0$ and   $V=V_c+\delta V$ in (\ref{eqn:delta2}). From (\ref{eqn:V2}) we get  $V(T)/V_c\approx 1-(\pi V_c\epsilon_F/6)(T/\epsilon_F)^2$, which when substituted in (\ref{eqn:delta2})
gives
\begin{equation}
\bar{\Delta}(T)^2=\bar{\Delta}(0)^2+\frac{\pi^2\epsilon_F^2}{3} \left(\frac{T}{\epsilon_F}\right) ^2\hspace{0.25cm}\textrm{as}\hspace{0.25cm}V\rightarrow V_c^+
\label{eqn:dplus}
\end{equation}
where $\bar{\Delta}(0)^2=(2\pi)(\epsilon_F/V_c)(\delta V/V_c)$ gives the standard mean-field exponent of $1/2$  at $T=0$ \cite{punnoose_TRV_XL}. 

(\textit{iii}) Finally, at $V=V_c$, since  $\bar{\Delta}(0)=0$ at the QCP, we get from (\ref{eqn:dplus}) that  $\bar{\Delta}(T)=\pi T/\sqrt{3}$; the linear $T$ dependence is  distinct from that on either side of the QCP.

To conclude, our  main observation from Eqs. (\ref{eqn:dminus}) and (\ref{eqn:dplus}) is that contrary to standard expectations,  the  intraband  excitations at finite $T$ has a stabilizing effect that tend to increase $\bar{\Delta}$ as $T$ increases. The increase will eventually be limited by the interband excitations when $T\gtrsim \epsilon_F$, the study of which is beyond the scope of this paper. More precisely, we note from Eq. (\ref{eqn:Tc}) that $T_c/\epsilon_F\sim \sqrt{(V_c-V)/\epsilon_F}$. Hence, provided  $\delta V \ll \epsilon_F$,  pair-breaking effects are parametrically far and our results are internally consistent with our approximations.

Finally, we want to stress the importance of studying the canonical free energy $\mathcal{F}(n_e)$ (since  the volume is held constant throughout, this translates to fixing the density), rather than directly  minimizing the action   $\tilde{\mathcal{L}}(\mu)$  holding $\mu$ constant. Although demanding $\partial \tilde{\mathcal{L}}/\partial \Delta=0$ recovers the saddle-point condition in  (\ref{eqn:saddle_point_lambda}), the condensate is unstable. To see this, we differentiate (\ref{eqn:Omega_lowT}) twice and obtain $\partial^2\tilde{\mathcal{L}}/\partial\Delta^2=-\Delta^2/\Lambda<0$. This instability was pointed out earlier by us in Ref. [\onlinecite{punnoose_TRV_XL}]  by an explicit diagrammatic calculation. In  Eq. (\ref{eqn:stability}) we resolve this instability by showing that the mean-field condensate obtained keeping $N$ fixed, rather than $\mu$, is indeed stable.

Notice that the order parameter preserves inversion ($\mathcal{I}$) symmetry but  breaks chiral ($\mathcal{C}$) symmetry in such a way that the combined $\mathcal{CT}$ symmetry is left invariant, as a result it can exhibit both anomalous Hall and Kerr effects \cite{haldane_TFL,fradkin_sun_TRV,honeycomb_spain_PRL}.  Furthermore, as noted in Ref.  [\onlinecite{fradkin_sun_TRV}], because of the chirality, the state does not couple  directly to non-magnetic impurities. Hence when spin effects can be ignored, the phase is expected to be stable. It remains to be seen, however, if the finite temperature phase described in this work survives in the presence of nearest-neighbor repulsion, which we have not included in this study.

Temperature induced topological transition has been reported  in the context of thermally induced band inversion due to electron-phonon interactions \cite{garate_PRL}. In our case, the topological state is induced by strong correlations and thus provides a qualitatively different paradigm involving only electronic degrees of freedom.

\begin{acknowledgements}
AP acknowledges E.\ Pont\'{o}n (IFT), J.\ Birman and M.\ C.\ N.\ Fiolhais  for helpful discussions,  and we thank P.\ Ghaemi  for drawing our attention to Ref.\,[\onlinecite{garate_PRL}].
\end{acknowledgements}

\section{appendix}


\section{[A1] Derivation of  the low temperature limit of $\mathcal{L}$}

In this section, we detail the steps taken to arrive at the low temperature limit for the action $\mathcal{L}$ derived in Eq. (8). We start from the definition 
\begin{equation*}
\mathcal{L}={\varphi^2}/{(2V)} +\mathcal{L}_{+} + \mathcal{L}_{-}
\end{equation*}
where $\mathcal{L}_\pm$ are defined in Eq. (4). Following the discussions leading up to Eq. (8), it can be shown that  the sum over $\bm{k}$ in $\mathcal{L}_\pm$ can be expressed as an integral with the appropriate cut-offs as
\begin{equation*}
\mathcal{L}_{\pm}=-2T\!\int_0^{p_\Lambda}\frac{pdp}{2\pi}\ln\!\left(1+e^{-(\pm\alpha\sqrt{p^2+\varphi_\lambda^2}+|\mu|)/T}\right)
\end{equation*}
where $\mu=-|\mu|$. Substituting
$z_\pm=(\alpha\sqrt{p^2+\varphi_\lambda^2}\pm |\mu|)/T$ brings the integrals into the more familiar form:
\begin{equation*}
\mathcal{L}_{\pm} =-\frac{T^2}{\pi\alpha^2}\int\limits_{(|\Delta_\lambda|\pm|\mu|)/T}^{(\Lambda_\lambda\pm|\mu|)/T}dz\, (T z_\pm \mp |\mu|)\ln(1+e^{\mp z_\pm})
\end{equation*}
The above integrals can be represented in terms of the polylogarithm functions $\textrm{Li}_2$ and $\textrm{Li}_3$.  (See Ref. [24] for the definitions and properties of the polylog functions.) Rescaling $\tilde{\mathcal{L}}=(\pi\alpha^2)\mathcal{L}$, we obtain
\begin{eqnarray*}
 \tilde{\mathcal{L}}_{+}&=&\frac{\Lambda_\lambda^3}{3}-\frac{|\Delta_\lambda|^3}{3}+\frac{\Lambda_\lambda^2|\mu|}{2}-\frac{\Delta_\lambda^2|\mu|}{2}\nonumber\\
&&+ T^2\left[\Lambda_\lambda \textrm{Li}_2\left(-e^{\frac{\Lambda_\lambda+|\mu|}{T}}\right)-|\Delta_\lambda| \textrm{Li}_2\left(-e^{\frac{|\Delta_\lambda|+|\mu|}{T}}\right)\right]\nonumber\\
&&-T^3\left[\textrm{Li}_3\left(-e^{\frac{\Lambda_\lambda+|\mu|}{T}}\right)
-\textrm{Li}_3\left(-e^{\frac{|\Delta_\lambda|+|\mu|}{T}}\right)\right]
\label{eqn:app:Omega+}\\
\tilde{\mathcal{L}}_{-}&=&T^2\left[\Lambda_\lambda\textrm{Li}_2\left(-e^{\frac{\Lambda_\lambda-|\mu|}{T}}\right)-|\Delta_\lambda| \textrm{Li}_2\left(-e^{\frac{|\Delta_\lambda|-|\mu|}{T}}\right)\right]\nonumber\\
&&-T^3\left[\textrm{Li}_3\left(-e^{\frac{\Lambda_\lambda-|\mu|}{T}}\right)
-\textrm{Li}_3\left(-e^{\frac{|\Delta_\lambda|-|\mu|}{T}}\right)\right]
\label{eqn:app:Omega-}
\end{eqnarray*}

To obtain the low temperature expansion, we use the  asymptotic properties of the  polylog functions. The required formulas to order $e^{-z}$ are listed below: (the ellipses indicate higher powers of $e^{-z}$)
\begin{eqnarray*}
\lim_{\textrm{Re}(z)\rightarrow \infty}
\textrm{Li}_2(-e^{-z})&=&-e^{-z}+\cdots\nonumber\\
\textrm{Li}_2(-e^z)&=&-\frac{z^2}{2}-\frac{\pi^2}{6}+e^{-z}+\cdots\nonumber\\
\textrm{Li}_3(-e^{-z})&=&-e^{-z}+\cdots \nonumber\\
\textrm{Li}_3(-e^z)&=&-\frac{z^3}{6}-\frac{\pi^2}{6}z-e^{-z}+\cdots
\end{eqnarray*}
Dropping terms that fall off as $e^{-|\mu|/T}$ and higher, we get $\mathcal{L}_+=0$. Note that  $|\Delta_\lambda|< E_{\lambda,F}=\sqrt{\Delta_\lambda^2+\epsilon_F^2}$ away from half-filling and $\Lambda_\lambda \gg \epsilon_\lambda$. Gathering the polynomial terms from $\tilde{\mathcal{L}}_-$ and substituting into  $\mathcal{L}$ gives Eq. (8).

\vspace{2\baselineskip}
\section{[A2] Derivation of the Equation of State}\label{app:EoS}
In this section, the algebraic steps leading up to the solution for the equation of state derived in (12) are detailed. This requires solving for $\lambda$ and $\mu$ using  the saddle-point equation (9) and the equation relating the chemical potential and density (10). 

The relevant equations to be solved are listed  below
\begin{eqnarray*}
\hspace{-3cm}\textrm{Eq. (9)}:\hspace{0.25cm} \frac{\pi\bar{\Delta}}{V} &=& \bar{\Delta}_\lambda\left(\bar{\Lambda}_\lambda-|\mu|\right)\\
\textrm{Eq.\,(10)}:\hspace{0.5cm} \mu^2 &=&2\tilde{n}_h+\bar{\Delta}_\lambda^2-\frac{\pi^2T^2}{3}
\end{eqnarray*}
where $\Lambda_\lambda=\sqrt{2\tilde{n}+{\bar{\Delta}}^2}$. Our aim is to eliminate the square-roots from $\bar{\Lambda}_\lambda$ and $\mu$ and obtain a polynomial equation that can be easily solved for $\bar{\Delta}_\lambda$. 
To this end, we multiply both sides of the first equation by $(\bar{\Lambda}_\lambda+|\mu|)$, and use the second equation to eliminate $\mu^2$. This simplifies the equations to
\begin{eqnarray*}
\bar{\Lambda}_\lambda+|\mu| &=& \left(\frac{V}{\pi\bar{\Delta}}\right) (2\tilde{n}-2\tilde{n}_h+\pi^2T^2/3)\bar{\Delta}_\lambda \\
\bar{\Lambda}_\lambda-|\mu| &=& \left(\frac{\pi\bar{\Delta}}{V}\right)\frac{1}{\bar{\Delta}_\lambda}
\end{eqnarray*}

It is now straightforward to solve for  $\bar{\Delta}_\lambda$ by adding  the two equations and squaring the result to eliminate the square-root. After some minimal algebra, we get
\begin{widetext}
\begin{equation*}
\bar{\Delta}_\lambda^2=(\bar{\Delta}+\alpha\lambda)^2=\left(\frac{\pi\bar{\Delta}}{V}\right)^2\left(
\frac{(2\tilde{n}+2\tilde{n}_h-\pi^2T^2/3)+2\sqrt{2\tilde{n}(2\tilde{n}_h-\pi^2T^2/3)+(\pi\bar{\Delta}/V)^2}}
{(2\tilde{n}-2\tilde{n}_h+\pi^2T^2/3)^2-4(\pi\bar{\Delta}/V)^2}\right)
\end{equation*}
\end{widetext}
Care must be taken in choosing the correct sign of the square-root in the numerator when solving the quadratic equation for $\bar{\Delta}_\lambda^2$. It is chosen to reproduce the solution for $V_c$ in Eq. (13), which can be independently obtained  from the saddle-point equation in the limit $\bar{\Delta}=\lambda=0$.

Finally, we take the positive root of $\bar{\Delta}_\lambda^2$ to obtain the equation of state given in (12). This is followed by substituting the solution in the equation for $\mu^2$  to find $|\mu|$. The solutions for $\lambda$ and $|\mu|$ are used to obtain the exact low-temperature canonical free energy $\mathcal{F}$, defined in Eq. (11). The result for $\delta\mathcal{F}$ is plotted in Fig. 1.


\begin{thebibliography}{33}%
\makeatletter
\providecommand \@ifxundefined [1]{%
 \@ifx{#1\undefined}
}%
\providecommand \@ifnum [1]{%
 \ifnum #1\expandafter \@firstoftwo
 \else \expandafter \@secondoftwo
 \fi
}%
\providecommand \@ifx [1]{%
 \ifx #1\expandafter \@firstoftwo
 \else \expandafter \@secondoftwo
 \fi
}%
\providecommand \natexlab [1]{#1}%
\providecommand \enquote  [1]{``#1''}%
\providecommand \bibnamefont  [1]{#1}%
\providecommand \bibfnamefont [1]{#1}%
\providecommand \citenamefont [1]{#1}%
\providecommand \href@noop [0]{\@secondoftwo}%
\providecommand \href [0]{\begingroup \@sanitize@url \@href}%
\providecommand \@href[1]{\@@startlink{#1}\@@href}%
\providecommand \@@href[1]{\endgroup#1\@@endlink}%
\providecommand \@sanitize@url [0]{\catcode `\\12\catcode `\$12\catcode
  `\&12\catcode `\#12\catcode `\^12\catcode `\_12\catcode `\%12\relax}%
\providecommand \@@startlink[1]{}%
\providecommand \@@endlink[0]{}%
\providecommand \url  [0]{\begingroup\@sanitize@url \@url }%
\providecommand \@url [1]{\endgroup\@href {#1}{\urlprefix }}%
\providecommand \urlprefix  [0]{URL }%
\providecommand \Eprint [0]{\href }%
\providecommand \doibase [0]{http://dx.doi.org/}%
\providecommand \selectlanguage [0]{\@gobble}%
\providecommand \bibinfo  [0]{\@secondoftwo}%
\providecommand \bibfield  [0]{\@secondoftwo}%
\providecommand \translation [1]{[#1]}%
\providecommand \BibitemOpen [0]{}%
\providecommand \bibitemStop [0]{}%
\providecommand \bibitemNoStop [0]{.\EOS\space}%
\providecommand \EOS [0]{\spacefactor3000\relax}%
\providecommand \BibitemShut  [1]{\csname bibitem#1\endcsname}%
\let\auto@bib@innerbib\@empty
\bibitem [{\citenamefont {Haldane}(1988)}]{haldane_graphene_TRV}%
  \BibitemOpen
  \bibfield  {author} {\bibinfo {author} {\bibfnamefont {F.~D.~M.}\
  \bibnamefont {Haldane}},\ }\href@noop {} {\bibfield  {journal} {\bibinfo
  {journal} {Phys. Rev. Lett.}\ }\textbf {\bibinfo {volume} {61}},\ \bibinfo
  {pages} {2015} (\bibinfo {year} {1988})}\BibitemShut {NoStop}%
\bibitem [{\citenamefont {Mott}(1961)}]{Mott_Transition}%
  \BibitemOpen
  \bibfield  {author} {\bibinfo {author} {\bibfnamefont {N.~F.}\ \bibnamefont
  {Mott}},\ }\href@noop {} {\bibfield  {journal} {\bibinfo  {journal} {Philos.
  Mag.}\ }\textbf {\bibinfo {volume} {6}},\ \bibinfo {pages} {287} (\bibinfo
  {year} {1961})}\BibitemShut {NoStop}%
\bibitem [{\citenamefont {Knox}(1963)}]{knox}%
  \BibitemOpen
  \bibfield  {author} {\bibinfo {author} {\bibfnamefont {R.~S.}\ \bibnamefont
  {Knox}},\ }\enquote {\bibinfo {title} {Theory of excitons},}\ in\ \href@noop
  {} {\emph {\bibinfo {booktitle} {Solid State Physics Suppl.}}},\
  Vol.~\bibinfo {volume} {5},\ \bibinfo {editor} {edited by\ \bibinfo {editor}
  {\bibfnamefont {F.}~\bibnamefont {Seitz}}\ and\ \bibinfo {editor}
  {\bibfnamefont {D.}~\bibnamefont {Turnbull}}}\ (\bibinfo  {publisher}
  {Academic Press, Inc., New York},\ \bibinfo {year} {1963})\ p.\ \bibinfo
  {pages} {100}\BibitemShut {NoStop}%
\bibitem [{\citenamefont {des Cloizeaux}(1965)}]{Cloizeaux}%
  \BibitemOpen
  \bibfield  {author} {\bibinfo {author} {\bibfnamefont {J.}~\bibnamefont {des
  Cloizeaux}},\ }\href@noop {} {\bibfield  {journal} {\bibinfo  {journal} {J.
  Phys. Chem. Solids}\ }\textbf {\bibinfo {volume} {26}},\ \bibinfo {pages}
  {259} (\bibinfo {year} {1965})}\BibitemShut {NoStop}%
\bibitem [{\citenamefont {Keldysh}\ and\ \citenamefont
  {Kopaev}(1965)}]{Keldysh_Kopaev}%
  \BibitemOpen
  \bibfield  {author} {\bibinfo {author} {\bibfnamefont {L.~V.}\ \bibnamefont
  {Keldysh}}\ and\ \bibinfo {author} {\bibfnamefont {Y.~V.}\ \bibnamefont
  {Kopaev}},\ }\href@noop {} {\bibfield  {journal} {\bibinfo  {journal} {Sov.
  Phys. Solid State}\ }\textbf {\bibinfo {volume} {6}},\ \bibinfo {pages}
  {2219} (\bibinfo {year} {1965})}\BibitemShut {NoStop}%
\bibitem [{\citenamefont {Kozlov}\ and\ \citenamefont
  {Maksimov}(1965)}]{Kozlov_Maksimov}%
  \BibitemOpen
  \bibfield  {author} {\bibinfo {author} {\bibfnamefont {A.~N.}\ \bibnamefont
  {Kozlov}}\ and\ \bibinfo {author} {\bibfnamefont {L.~A.}\ \bibnamefont
  {Maksimov}},\ }\href@noop {} {\bibfield  {journal} {\bibinfo  {journal} {Sov.
  Phys. JETP}\ }\textbf {\bibinfo {volume} {21}},\ \bibinfo {pages} {790}
  (\bibinfo {year} {1965})}\BibitemShut {NoStop}%
\bibitem [{\citenamefont {Jerome}\ \emph {et~al.}(1967)\citenamefont {Jerome},
  \citenamefont {Rice},\ and\ \citenamefont {Kohn}}]{jerome_rice_kohn}%
  \BibitemOpen
  \bibfield  {author} {\bibinfo {author} {\bibfnamefont {D.}~\bibnamefont
  {Jerome}}, \bibinfo {author} {\bibfnamefont {T.~M.}\ \bibnamefont {Rice}}, \
  and\ \bibinfo {author} {\bibfnamefont {W.}~\bibnamefont {Kohn}},\ }\href@noop
  {} {\bibfield  {journal} {\bibinfo  {journal} {Phys. Rev.}\ }\textbf
  {\bibinfo {volume} {158}},\ \bibinfo {pages} {462} (\bibinfo {year}
  {1967})}\BibitemShut {NoStop}%
\bibitem [{\citenamefont {Halperin}\ and\ \citenamefont
  {Rice}(1968)}]{RMP_halperin_rice}%
  \BibitemOpen
  \bibfield  {author} {\bibinfo {author} {\bibfnamefont {B.~I.}\ \bibnamefont
  {Halperin}}\ and\ \bibinfo {author} {\bibfnamefont {T.~M.}\ \bibnamefont
  {Rice}},\ }\href@noop {} {\bibfield  {journal} {\bibinfo  {journal} {Rev.
  Mod. Phys.}\ }\textbf {\bibinfo {volume} {40}},\ \bibinfo {pages} {755}
  (\bibinfo {year} {1968})}\BibitemShut {NoStop}%
\bibitem [{\citenamefont {Comte}\ and\ \citenamefont
  {Nozi\`{e}res}(1982)}]{Nozieres_Comte}%
  \BibitemOpen
  \bibfield  {author} {\bibinfo {author} {\bibfnamefont {C.}~\bibnamefont
  {Comte}}\ and\ \bibinfo {author} {\bibfnamefont {P.}~\bibnamefont
  {Nozi\`{e}res}},\ }\href@noop {} {\bibfield  {journal} {\bibinfo  {journal}
  {J. Phys. (Paris)}\ }\textbf {\bibinfo {volume} {43}},\ \bibinfo {pages}
  {1069} (\bibinfo {year} {1982})}\BibitemShut {NoStop}%
\bibitem [{\citenamefont {Varma}(1997)}]{varma_1997}%
  \BibitemOpen
  \bibfield  {author} {\bibinfo {author} {\bibfnamefont {C.~M.}\ \bibnamefont
  {Varma}},\ }\href@noop {} {\bibfield  {journal} {\bibinfo  {journal} {Phys.
  Rev. B}\ }\textbf {\bibinfo {volume} {55}},\ \bibinfo {pages} {14554}
  (\bibinfo {year} {1997})}\BibitemShut {NoStop}%
\bibitem [{\citenamefont {Sun}\ and\ \citenamefont
  {Fradkin}(2008)}]{fradkin_sun_TRV}%
  \BibitemOpen
  \bibfield  {author} {\bibinfo {author} {\bibfnamefont {K.}~\bibnamefont
  {Sun}}\ and\ \bibinfo {author} {\bibfnamefont {E.}~\bibnamefont {Fradkin}},\
  }\href@noop {} {\bibfield  {journal} {\bibinfo  {journal} {Phys. Rev. B}\
  }\textbf {\bibinfo {volume} {78}},\ \bibinfo {pages} {245122} (\bibinfo
  {year} {2008})}\BibitemShut {NoStop}%
\bibitem [{\citenamefont {Liu}\ and\ \citenamefont
  {Punnoose}(2014)}]{punnoose_TRV_XL}%
  \BibitemOpen
  \bibfield  {author} {\bibinfo {author} {\bibfnamefont {W.}~\bibnamefont
  {Liu}}\ and\ \bibinfo {author} {\bibfnamefont {A.}~\bibnamefont {Punnoose}},\
  }\href@noop {} {\bibfield  {journal} {\bibinfo  {journal} {Phys. Rev. B}\
  }\textbf {\bibinfo {volume} {89}},\ \bibinfo {pages} {045126} (\bibinfo
  {year} {2014})}\BibitemShut {NoStop}%
\bibitem [{\citenamefont {Haldane}(2004)}]{haldane_TFL}%
  \BibitemOpen
  \bibfield  {author} {\bibinfo {author} {\bibfnamefont {F.~D.~M.}\
  \bibnamefont {Haldane}},\ }\href@noop {} {\bibfield  {journal} {\bibinfo
  {journal} {Phys. Rev. Lett.}\ }\textbf {\bibinfo {volume} {93}},\ \bibinfo
  {pages} {206602} (\bibinfo {year} {2004})}\BibitemShut {NoStop}%
\bibitem [{\citenamefont {Varma}\ \emph {et~al.}(2002)\citenamefont {Varma},
  \citenamefont {Nussinov},\ and\ \citenamefont {van
  Saarloos}}]{varma_SFL_review}%
  \BibitemOpen
  \bibfield  {author} {\bibinfo {author} {\bibfnamefont {C.~M.}\ \bibnamefont
  {Varma}}, \bibinfo {author} {\bibfnamefont {Z.}~\bibnamefont {Nussinov}}, \
  and\ \bibinfo {author} {\bibfnamefont {W.}~\bibnamefont {van Saarloos}},\
  }\href@noop {} {\bibfield  {journal} {\bibinfo  {journal} {Phys. Rep.}\
  }\textbf {\bibinfo {volume} {361}},\ \bibinfo {pages} {267} (\bibinfo {year}
  {2002})}\BibitemShut {NoStop}%
\bibitem [{\citenamefont {Castro}\ \emph {et~al.}(2011)\citenamefont {Castro},
  \citenamefont {Grushin}, \citenamefont {Valenzuela}, \citenamefont
  {Vozmediano}, \citenamefont {Cortijo},\ and\ \citenamefont
  {de~Juan}}]{honeycomb_spain_PRL}%
  \BibitemOpen
  \bibfield  {author} {\bibinfo {author} {\bibfnamefont {E.~V.}\ \bibnamefont
  {Castro}}, \bibinfo {author} {\bibfnamefont {A.~G.}\ \bibnamefont {Grushin}},
  \bibinfo {author} {\bibfnamefont {B.}~\bibnamefont {Valenzuela}}, \bibinfo
  {author} {\bibfnamefont {M.~A.~H.}\ \bibnamefont {Vozmediano}}, \bibinfo
  {author} {\bibfnamefont {A.}~\bibnamefont {Cortijo}}, \ and\ \bibinfo
  {author} {\bibfnamefont {F.}~\bibnamefont {de~Juan}},\ }\href@noop {}
  {\bibfield  {journal} {\bibinfo  {journal} {Phys. Rev. Lett.}\ }\textbf
  {\bibinfo {volume} {107}},\ \bibinfo {pages} {106402} (\bibinfo {year}
  {2011})}\BibitemShut {NoStop}%
\bibitem [{\citenamefont {Raghu}\ \emph {et~al.}(2008)\citenamefont {Raghu},
  \citenamefont {Qi}, \citenamefont {Honerkamp},\ and\ \citenamefont
  {Zhang}}]{Zhang_TMI}%
  \BibitemOpen
  \bibfield  {author} {\bibinfo {author} {\bibfnamefont {S.}~\bibnamefont
  {Raghu}}, \bibinfo {author} {\bibfnamefont {X.-L.}\ \bibnamefont {Qi}},
  \bibinfo {author} {\bibfnamefont {C.}~\bibnamefont {Honerkamp}}, \ and\
  \bibinfo {author} {\bibfnamefont {S.-C.}\ \bibnamefont {Zhang}},\ }\href@noop
  {} {\bibfield  {journal} {\bibinfo  {journal} {Phys. Rev. Lett.}\ }\textbf
  {\bibinfo {volume} {100}},\ \bibinfo {pages} {156401} (\bibinfo {year}
  {2008})}\BibitemShut {NoStop}%
\bibitem [{\citenamefont {Sun}\ \emph {et~al.}(2009)\citenamefont {Sun},
  \citenamefont {Yao}, \citenamefont {Fradkin},\ and\ \citenamefont
  {Kivelson}}]{QBCP_fradkin}%
  \BibitemOpen
  \bibfield  {author} {\bibinfo {author} {\bibfnamefont {K.}~\bibnamefont
  {Sun}}, \bibinfo {author} {\bibfnamefont {H.}~\bibnamefont {Yao}}, \bibinfo
  {author} {\bibfnamefont {E.}~\bibnamefont {Fradkin}}, \ and\ \bibinfo
  {author} {\bibfnamefont {S.~A.}\ \bibnamefont {Kivelson}},\ }\href@noop {}
  {\bibfield  {journal} {\bibinfo  {journal} {Phys. Rev. Lett.}\ }\textbf
  {\bibinfo {volume} {103}},\ \bibinfo {pages} {046811} (\bibinfo {year}
  {2009})}\BibitemShut {NoStop}%
\bibitem [{\citenamefont {Weeks}\ and\ \citenamefont
  {Franz}(2010)}]{honeycomb_Weeks_Franz}%
  \BibitemOpen
  \bibfield  {author} {\bibinfo {author} {\bibfnamefont {C.}~\bibnamefont
  {Weeks}}\ and\ \bibinfo {author} {\bibfnamefont {M.}~\bibnamefont {Franz}},\
  }\href@noop {} {\bibfield  {journal} {\bibinfo  {journal} {Phys. Rev. B}\
  }\textbf {\bibinfo {volume} {81}},\ \bibinfo {pages} {085105} (\bibinfo
  {year} {2010})}\BibitemShut {NoStop}%
\bibitem [{\citenamefont {Wen}\ \emph {et~al.}(2010)\citenamefont {Wen},
  \citenamefont {R\"{u}egg}, \citenamefont {Wang},\ and\ \citenamefont
  {Fiete}}]{honeycomb_fiete}%
  \BibitemOpen
  \bibfield  {author} {\bibinfo {author} {\bibfnamefont {J.}~\bibnamefont
  {Wen}}, \bibinfo {author} {\bibfnamefont {A.}~\bibnamefont {R\"{u}egg}},
  \bibinfo {author} {\bibfnamefont {C.-C.~J.}\ \bibnamefont {Wang}}, \ and\
  \bibinfo {author} {\bibfnamefont {G.~A.}\ \bibnamefont {Fiete}},\ }\href@noop
  {} {\bibfield  {journal} {\bibinfo  {journal} {Phys. Rev. B}\ }\textbf
  {\bibinfo {volume} {82}},\ \bibinfo {pages} {075125} (\bibinfo {year}
  {2010})}\BibitemShut {NoStop}%
\bibitem [{\citenamefont {Tieleman}\ \emph {et~al.}(2013)\citenamefont
  {Tieleman}, \citenamefont {Dutta}, \citenamefont {Lewenstein},\ and\
  \citenamefont {Eckardt}}]{triangle_tieleman}%
  \BibitemOpen
  \bibfield  {author} {\bibinfo {author} {\bibfnamefont {O.}~\bibnamefont
  {Tieleman}}, \bibinfo {author} {\bibfnamefont {O.}~\bibnamefont {Dutta}},
  \bibinfo {author} {\bibfnamefont {M.}~\bibnamefont {Lewenstein}}, \ and\
  \bibinfo {author} {\bibfnamefont {A.}~\bibnamefont {Eckardt}},\ }\href@noop
  {} {\bibfield  {journal} {\bibinfo  {journal} {Phys. Rev. Lett.}\ }\textbf
  {\bibinfo {volume} {110}},\ \bibinfo {pages} {096405} (\bibinfo {year}
  {2013})}\BibitemShut {NoStop}%
\bibitem [{\citenamefont {Kurita}\ \emph {et~al.}(2013)\citenamefont {Kurita},
  \citenamefont {Yamaji},\ and\ \citenamefont {Imada}}]{imada_gapless}%
  \BibitemOpen
  \bibfield  {author} {\bibinfo {author} {\bibfnamefont {M.}~\bibnamefont
  {Kurita}}, \bibinfo {author} {\bibfnamefont {Y.}~\bibnamefont {Yamaji}}, \
  and\ \bibinfo {author} {\bibfnamefont {M.}~\bibnamefont {Imada}},\
  }\href@noop {} {\bibfield  {journal} {\bibinfo  {journal} {Phys. Rev. B}\
  }\textbf {\bibinfo {volume} {88}},\ \bibinfo {pages} {115143} (\bibinfo
  {year} {2013})}\BibitemShut {NoStop}%
\bibitem [{\citenamefont {Grushin}\ \emph {et~al.}(2013)\citenamefont
  {Grushin}, \citenamefont {Castro}, \citenamefont {Cortijo}, \citenamefont
  {de~Juan}, \citenamefont {Vozmediano},\ and\ \citenamefont
  {Valenzuela}}]{honeycomb_spain_PRB}%
  \BibitemOpen
  \bibfield  {author} {\bibinfo {author} {\bibfnamefont {A.~G.}\ \bibnamefont
  {Grushin}}, \bibinfo {author} {\bibfnamefont {E.~V.}\ \bibnamefont {Castro}},
  \bibinfo {author} {\bibfnamefont {A.}~\bibnamefont {Cortijo}}, \bibinfo
  {author} {\bibfnamefont {F.}~\bibnamefont {de~Juan}}, \bibinfo {author}
  {\bibfnamefont {M.~A.~H.}\ \bibnamefont {Vozmediano}}, \ and\ \bibinfo
  {author} {\bibfnamefont {B.}~\bibnamefont {Valenzuela}},\ }\href@noop {}
  {\bibfield  {journal} {\bibinfo  {journal} {Phys. Rev. B}\ }\textbf {\bibinfo
  {volume} {87}},\ \bibinfo {pages} {085136} (\bibinfo {year}
  {2013})}\BibitemShut {NoStop}%
\bibitem [{\citenamefont {Duri\'{c}}\ \emph {et~al.}(2014)\citenamefont
  {Duri\'{c}}, \citenamefont {Chancellor},\ and\ \citenamefont
  {Herbut}}]{honeycomb_herbut}%
  \BibitemOpen
  \bibfield  {author} {\bibinfo {author} {\bibfnamefont {T.}~\bibnamefont
  {Duri\'{c}}}, \bibinfo {author} {\bibfnamefont {N.}~\bibnamefont
  {Chancellor}}, \ and\ \bibinfo {author} {\bibfnamefont {I.~F.}\ \bibnamefont
  {Herbut}},\ }\href@noop {} {\bibfield  {journal} {\bibinfo  {journal} {Phys.
  Rev. B}\ }\textbf {\bibinfo {volume} {89}},\ \bibinfo {pages} {165123}
  (\bibinfo {year} {2014})}\BibitemShut {NoStop}%
\bibitem [{\citenamefont {Tinkham}(1992)}]{tinkham_book}%
  \BibitemOpen
  \bibfield  {author} {\bibinfo {author} {\bibfnamefont {M.}~\bibnamefont
  {Tinkham}},\ }\href@noop {} {\emph {\bibinfo {title} {Group Theory and
  Quantum Mechanics}}}\ (\bibinfo  {publisher} {Dover Publications, New York},\
  \bibinfo {year} {1992})\BibitemShut {NoStop}%
\bibitem [{\citenamefont {de~Gennes}(1963)}]{deGennes_Ising}%
  \BibitemOpen
  \bibfield  {author} {\bibinfo {author} {\bibfnamefont {P.~G.}\ \bibnamefont
  {de~Gennes}},\ }\href@noop {} {\bibfield  {journal} {\bibinfo  {journal}
  {Solid State Commun.}\ }\textbf {\bibinfo {volume} {1}},\ \bibinfo {pages}
  {132} (\bibinfo {year} {1963})}\BibitemShut {NoStop}%
\bibitem [{\citenamefont {Brout}\ \emph {et~al.}(1966)\citenamefont {Brout},
  \citenamefont {M\"{u}ller},\ and\ \citenamefont {Thomas}}]{brout_Ising}%
  \BibitemOpen
  \bibfield  {author} {\bibinfo {author} {\bibfnamefont {R.}~\bibnamefont
  {Brout}}, \bibinfo {author} {\bibfnamefont {K.~A.}\ \bibnamefont
  {M\"{u}ller}}, \ and\ \bibinfo {author} {\bibfnamefont {H.}~\bibnamefont
  {Thomas}},\ }\href@noop {} {\bibfield  {journal} {\bibinfo  {journal} {Solid
  State Commun.}\ }\textbf {\bibinfo {volume} {4}},\ \bibinfo {pages} {507}
  (\bibinfo {year} {1966})}\BibitemShut {NoStop}%
\bibitem [{\citenamefont {Imada}\ \emph {et~al.}(2010)\citenamefont {Imada},
  \citenamefont {Misawa},\ and\ \citenamefont {Yamaji}}]{imada_review}%
  \BibitemOpen
  \bibfield  {author} {\bibinfo {author} {\bibfnamefont {M.}~\bibnamefont
  {Imada}}, \bibinfo {author} {\bibfnamefont {T.}~\bibnamefont {Misawa}}, \
  and\ \bibinfo {author} {\bibfnamefont {Y.}~\bibnamefont {Yamaji}},\
  }\href@noop {} {\bibfield  {journal} {\bibinfo  {journal} {J. Phys.: Condens.
  Matter}\ }\textbf {\bibinfo {volume} {22}},\ \bibinfo {pages} {164206}
  (\bibinfo {year} {2010})}\BibitemShut {NoStop}%
\bibitem [{\citenamefont {Altland}\ and\ \citenamefont
  {Simons}(2010)}]{book_Altland_Simons}%
  \BibitemOpen
  \bibfield  {author} {\bibinfo {author} {\bibfnamefont {A.}~\bibnamefont
  {Altland}}\ and\ \bibinfo {author} {\bibfnamefont {B.}~\bibnamefont
  {Simons}},\ }\href@noop {} {\emph {\bibinfo {title} {Condensed Matter Field
  Theory}}},\ \bibinfo {edition} {2nd}\ ed.\ (\bibinfo  {publisher} {Cambridge
  Univeristy Press},\ \bibinfo {year} {2010})\BibitemShut {NoStop}%
\bibitem [{\citenamefont {Abrikosov}\ \emph {et~al.}(1975)\citenamefont
  {Abrikosov}, \citenamefont {Gorkov},\ and\ \citenamefont
  {Dzyaloshinski}}]{bluebook}%
  \BibitemOpen
  \bibfield  {author} {\bibinfo {author} {\bibfnamefont {A.~A.}\ \bibnamefont
  {Abrikosov}}, \bibinfo {author} {\bibfnamefont {L.~P.}\ \bibnamefont
  {Gorkov}}, \ and\ \bibinfo {author} {\bibfnamefont {I.~E.}\ \bibnamefont
  {Dzyaloshinski}},\ }\href@noop {} {\emph {\bibinfo {title} {Methods of
  Quantum Field Theory in Statistical Physics}}}\ (\bibinfo  {publisher} {Dover
  Publications, New York},\ \bibinfo {year} {1975})\BibitemShut {NoStop}%
\bibitem [{Note1()}]{Note1}%
  \BibitemOpen
  \bibinfo {note} {The TMI has been analyzed in detail in Ref. [\protect
  \rev@citealp {imada_gapless}]; the deviation from the standard LGW theory,
  peculiar to the half-filled case at $T=0$, vanish both at finite $T$ and/or
  doping. For the interested reader, the relationship between the parameters in
  [\protect \rev@citealp {imada_gapless}] with those in this paper are: $\zeta
  =\varphi /V,\protect \tmspace +\thinmuskip {.1667em} \Lambda =p_\Lambda
  ,\protect \tmspace +\thinmuskip {.1667em} t=\alpha /\protect \sqrt {3}$ and $
  V_2/t=V/3$.}\BibitemShut {Stop}%
\bibitem [{\citenamefont {Abramowitz}\ and\ \citenamefont
  {Stegun}(1964)}]{book_functions}%
  \BibitemOpen
  \bibinfo {editor} {\bibfnamefont {M.}~\bibnamefont {Abramowitz}}\ and\
  \bibinfo {editor} {\bibfnamefont {I.~A.}\ \bibnamefont {Stegun}},\ eds.,\
  \href@noop {} {\emph {\bibinfo {title} {Handbook of Mathematical
  Functions}}}\ (\bibinfo  {publisher} {National Bureau of Standards,
  Washington, DC},\ \bibinfo {year} {1964})\BibitemShut {NoStop}%
\bibitem [{\citenamefont {Negele}\ and\ \citenamefont
  {Orland}(1998)}]{book_negele_orland}%
  \BibitemOpen
  \bibfield  {author} {\bibinfo {author} {\bibfnamefont {J.~W.}\ \bibnamefont
  {Negele}}\ and\ \bibinfo {author} {\bibfnamefont {H.}~\bibnamefont
  {Orland}},\ }\href@noop {} {\emph {\bibinfo {title} {Quantum many-particle
  systems}}}\ (\bibinfo  {publisher} {Perseus Books, Massachusetts},\ \bibinfo
  {year} {1998})\BibitemShut {NoStop}%
\bibitem [{\citenamefont {Garate}(2013)}]{garate_PRL}%
  \BibitemOpen
  \bibfield  {author} {\bibinfo {author} {\bibfnamefont {I.}~\bibnamefont
  {Garate}},\ }\href@noop {} {\bibfield  {journal} {\bibinfo  {journal} {Phys.
  Rev. Lett.}\ }\textbf {\bibinfo {volume} {110}},\ \bibinfo {pages} {046402}
  (\bibinfo {year} {2013})}\BibitemShut {NoStop}%
\end{thebibliography}
%

\end{document}